\documentclass[article,twocolumn,superscriptaddress,amsmath,amssymb,aps,nofootinbib]{revtex4-2}
\usepackage[utf8]{inputenc}
\usepackage{amsmath,amsthm,amssymb,amsfonts,braket}
\usepackage{placeins}[verbose]
\usepackage{tabularx, longtable, multirow}
\usepackage{graphicx}
\usepackage{gensymb} 
\usepackage{dcolumn}
\usepackage{bm,dsfont}
\usepackage[english]{babel}
\usepackage{comment}
\usepackage[dvipsnames]{xcolor}
\usepackage{siunitx}
\usepackage[normalem]{ulem}
\usepackage[T1]{fontenc}
\usepackage{physics}
\usepackage{mathtools}
\usepackage{float}
\usepackage{hyperref}

\begin{document}

\title{Planckian Bounds From Local Uncertainty Relations}

\author{Zohar Nussinov}
\email{zohar@wustl.edu}
\affiliation{
Department of Physics, Washington University in St. Louis, 1 Brookings Drive, St. Louis, MO 63130,
  USA}
\affiliation{Institut für Physik, Technische Universität Chemnitz, 09107 Chemnitz, Germany
}
\author{Saurish Chakrabarty}
\email{saurish@apccollege.ac.in}
\affiliation{
Department of Physics, Acharya Prafulla Chandra College, New Barrackpore, Kolkata 700131, India}

\date{\today}

\begin{abstract}
  We introduce ``local uncertainty relations'' in thermal many-body systems, from which fundamental bounds in quantum systems can be derived. These lead to universal non-relativistic speed limits (independent of interaction range) and transport coefficient bounds (e.g., those of the diffusion constant and viscosity)  that are compared against experimental data. 

\end{abstract}

\maketitle
\section{Personal note concerning Jan Zaanen and the current work}
It is an honor combined with great sadness to write in this volume dedicated to Jan Zaanen. As those who knew him could readily attest, Jan was a very engaging and deep minded person with a strong passion for science and much else. Exciting conversations with Jan could go on for many hours and span nearly any topic. One of us (ZN) had the privilege of being Jan's postdoc. Jan was exceedingly generous with his ideas and time and apart from providing many of his nsights often also encouraged his group members to pursue their own interests using himself as a sounding board. 
Just as ZN arrived at Leiden as a his postdoc nearly 25 years ago, Jan already propelled (especially  at that time) daring suggestions regarding connections between gravity and strongly correlated electronic systems (giving rise, en passant, to works on his ideas on ``quantum elasticity''). Jan was also already then fascinated with various aspects of topological orders (including, in particular, non-local ``string order'' correlation functions) in electronic systems. He loved dualities.  Jan was not confined to a specific niche but rather continuously kept moving into new terrain in the pursuit of trying to fundamentally understand things. Although some of Jan's ideas seemed far too revolutionary when first hearing them- in the end Jan's intuition invariably proved to be correct or lead to questions/directions that were incredibly rich and often well ahead of their time. Combined with his passion for things very theoretical, Jan also possessed an excellent intuitive knack for rationalizing experimental results. The ``stripe club'' meetings of Jan's group and later on also other groups at the Instituut Lorentz led to lively discussions that triggered many insights. Given that ``Planckian dissipation'' \cite{zaanen2004} was one of the last areas that we interacted with Jan before his untimely illness, we thought only appropriate to revisit results related to these and their comparison to experiment. Our aim here is not only to do so with strongly correlated electronic systems in mind but also consider more mundane seemingly ``classical'' liquids and gases. We illustrate how Planck's constant sets rigorous bounds on spatial gradients of general observables and their time derivatives in both thermal and athermal systems implying corollaries on transport coefficients.

\section{Introduction}
The prominent appearance of Planck's constant in ``classical'' textbook type systems is hardly new- Sackur and Tetrode's calculations \cite{Sackur,tetrode} for the entropy of ideal monatomic gases enabled early estimates of Planck's constant from thermodynamic measurements of mercury vapor \cite{experiment}. Our interest here is in illustrating that rigorous bounds (and simplified estimates of these bounds) on relaxation rates (and spatial gradients) in these and other general systems. Towards this end, in the current work, we will first briefly review principle inequalities that we derived in Refs.  \cite{planckianAOP,macroscopic} and then turn to contrasting these with experimental data.
Inspired by a recent illuminating preprint by Zhang {\it et al.} \cite{Heller}, a brief extension of our earlier bounds to disordered systems will be discussed in the Appendix. 
Before proceeding any further regarding our inequalities, we must pay tribute to numerous conjectured ``Planckian'' bounds which we will later return to. These include ``chaos bounds''
(i.e., those on the Lyapunov exponent) \cite{mssChaos},
bounds on lifetimes and relaxation rates, e.g., \cite{phil,planckianBruin,nnbk,Ramshaw,melting-speed,dassarma}, the viscosity \cite{Anup,nnbk,zoharViscosityPRR,jan-planck,trachenko2020viscositySciAdv,Lucas1},
the ratio of shear viscosity to entropy density \cite{KSS2}, the speed of sound and other speeds \cite{melting-speed,kostya2,Kostyabook}. Prominent works by Eyring \cite{Eyring1,Eyring2} and other pioneers examining chemical reaction rates raised the possibility of related inequalities. Here, our goal is to explain how such bounds result from local uncertainty relations in general quantum many body systems \cite{planckianAOP,macroscopic} and non-rigorous variants resulting from simple physical considerations. 

\section{The idea in a nutshell}
\label{nut}

The recipe underlying our bounds is an exceptionally simple one \cite{planckianAOP,macroscopic}. It combines two main ingredients:   \\

{\bf (1)} Our bounds can easily be obtained using the bare (open system variance type) quantum mechanical uncertainty relations between two operators in which one of the operators is chosen to be the system Hamiltonian or momentum. Importantly, even in systems harboring long range interactions, only a {\it few } terms may give rise to non-vanishing contributions to the relevant commutators. This leads to ``local'' uncertainty relations for typical few body observables of experimental interest. \\

{\bf (2)} The variances that appear in the resulting local uncertainty relations can, in many cases, often be determined precisely in the classical limit.  \\

Typically, a factor of Planck's constant originates from {\bf (1)} while (for a thermal system at a temperature $T$) energy scales ($k_{B} T)$ that are dimensionless multiples of $(k_{B} T)$ result from the evaluation of the variances in {\bf (2)}. Time energy uncertainty relations then {\it rigorously} yield minimal relaxation times for otherwise classical systems that scale as \footnote{The ratio appearing on the rightand side of Eq. (\ref{eq:time}) defines the ``Planckian'' time \cite{zaanen2004}. This prominent time scale appears in myriad contexts including, crucially, viable quantum critical phenomena \cite{Hertz,subir,marel,NFL}.}
\begin{eqnarray}
\label{eq:time}
\tau \ge {\cal{O}} \Big(\frac{\hbar}{k_{B} T} \Big).
\end{eqnarray}

Similarly, in a system of particles of mass $m$, the position-momentum uncertainly relations lead to bounds on the maximal gradients of general local observables in thermal equilibrium. The length scale $\lambda$ over which these vary appreciably (set by their relative gradient),
\begin{eqnarray}
\label{eq:space}
\lambda \ge {\cal{O}}(\lambda_{T}),
\end{eqnarray}
where, in non-relativistic systems, the thermal de-Broglie length $\lambda_{T} = \sqrt{2 \pi \hbar^2/(m k_{B} T)}$. 
In the classical ($\hbar \to 0$) limit, the lower bounds in both Eqs. (\ref{eq:time}, \ref{eq:space}) vanish. However, in the quantum arena, these inequalities impose genuine bounds in thermal systems and others that we will briefly discuss. 
When the operators at hand include, e.g., the electric potential, the inequality of Eq. (\ref{eq:space}) in thermal quantum systems, leads to bounds, on possible ratios between electric {\it field strengths} to the potentials of which they are a gradient. In the current work, we will 
largely examine lowest order corrections in $\hbar$ wherein the expectation values in the exact quantum inequalities will be evaluated classically.  \\

For completeness, we remark that an alternative 
analysis bypasses the use of the uncertainty relations of {\bf{(1)}}. Instead,  \\

{\bf(3)} Moments of spatial or temporal derivatives of general local observables that are, respectively, expressed as factors of $\hbar^{-1}$ (whose number depends on the order of the moment examined) times expectation values of the corresponding (moment order) power of the commutator of the local observable with the momentum (for spatial gradients) or non-commuting terms in the Hamiltonian (for time derivatives). These moments of the commutators are then simply written longhand as sums/differences of expectation values of individual operator string products that appear when multiplying the commutators. Up to literal counting factors, the resulting expectation value can then be exactly (and trivially) bounded by the single operator string product expectation value of the largest absolute value. In the classical limit, the associated expectation value- a phase space integral- may then be evaluated (or bounded). As may be expected by dimensional analysis and is indeed borne out by explicit calculation, this invariably similarly leads  \cite{planckianAOP} to Eqs. (\ref{eq:time}, \ref{eq:space}). Eq. (\ref{eq:space}) is derived along these lines for few-body observables that are a functions of the spatial coordinates alone (i.e., contain no momentum dependence). 

\section{Local uncertainty relations in open many body systems}
\label{Sec:local}
 The simple  (variance type) pure state quantum mechanical uncertainty relations  
\begin{eqnarray}
   \sigma_A~\sigma_B\ge\frac{1}{2}\left|\big\langle[A,B]\big\rangle\right|,
   \label{uncertainty}
 \end{eqnarray}
undergo no change \footnote{This extension is, of course, very well known yet not always derived  expediently. To quickly demonstrate this, we may, e.g., introduce \cite{planckianAOP}  a (modified trace type) inner product $(A,B) \equiv {\sf Tr}(\rho_{\Lambda} A^{\dagger}B)$ which is manifestly positive semi-definite, $(A,A) = {\sf Tr} (\rho_{\Lambda} A^{\dagger} A) \ge 0$ (and thus defines an inner product space for which the Cauchy–Schwarz inequality applies). We may now observe that the standard proof of the pure state variance type uncertainty relations can be repeated verbatim with the substitution of this inner product instead of the conventional one ($\langle a| b \rangle$) between states to immediately yield Eq. (\ref{uncertainty}). An alternate very short proof employs ancillas to trivially express general mixed stat averages in terms of those with pure states on a larger space  \cite{planckianAOP}.} when extended to arbitrary mixed states defined by the probability density $\rho_{\Lambda}$ of a general open many body system $\Lambda$. (In this latter case, the expectation value and standard deviations in Eq. (\ref{uncertainty}) are computed with $\rho_{\Lambda}$). If the operator $A$ is chosen to be the 
Hamiltonian of a general non-relativistic system $\Lambda$ that contains two-body interactions, 
\begin{eqnarray}
\label{eq:HLAMBDA}
H_{\Lambda}=\sum_{i=1}^{N_{\Lambda}} \frac{{\bf p}_{i}^{2}}{2 m}+\frac{1}{2} \sum_{i, j} {\cal{V}} \left({\bf r}_{i}, {\bf r}_{j}\right) \equiv K + V,
\end{eqnarray}
then, by Heisenberg's equations of motion, the righthand side of Eq. (\ref{uncertainty})
will correspond to the magnitude of the time derivative of an observable $B$. The variance of the many body Hamiltonian is extensive (typically scaling as $N^{1/2}$ for an open system of $N$ particles). Thus the resulting upper bound (lefthand side of Eq. (\ref{uncertainty})) diverges and hence is meaningless in the thermodynamic $N \to \infty$ limit. Now here is the exceptionally trivial yet important earlier point {\bf (1)}: when evaluating the commutator between $H_{\Lambda}$ and a general local operator $Q_{i}$ only a finite number of terms may be finite. Thus, only these terms (whose sum is a {\it local operator} $\tilde{H}_{i} \subset H_{\Lambda}$) may be kept. In what follows, we will largely work in the Heisenberg picture and often explicitly denote the time dependent operators in this picture by a superscript $^H$. Setting the operators in Eq. (\ref{uncertainty}) to be the Heisenberg picture $A=\tilde{H}_{i}^H$ and $B= Q_{i}^{H}$ may then lead to meaningful, system size independent, bounds on the time derivatives of general local observables. For instance, if the local observable is a general function of the $\ell$-th Cartesian position coordinate of the $i-$th particle ($1 \le i \le N$), i.e., $Q_{i}^{H} = f(r^{H}_{i \ell})$ then the only term in the Hamiltonian $H_{\Lambda}$ of Eq. (\ref{eq:HLAMBDA}) that will not commute with $f$ will be the {\it single body kinetic energy contribution} $\tilde{H}^{H}_{i}= (p^H_{i \ell})^{2}/(2m)$. Regardless of the form of the interaction $V$, the classical phase space variance (and higher order moments) of the $\tilde{H}^{H}_{i}= (p^H_{i \ell})^{2}/(2m)$ are trivial to compute for the Gaussian momentum distribution associated with the classically (momentum and coordinate) factorizable phase space probability density $\rho_{\Lambda}^{\sf classical~canonical} = 
e^{-\beta K} e^{-\beta V}/Z^{(cl)}_{\Lambda} $ (with the classical partition function $Z^{(cl)}_{\Lambda}$ 
of the open classical thermal system).  These will lead to bounds on the time derivative of $f$ (and higher order moments of the time derivative) of $f$ establishing the minimal time scale of Eq. (\ref{eq:time}) that are independent of the interactions ${\cal V}$ (including whether or not these are of a finite spatial range). For systems with bounded standard deviation of the particle displacements (e.g., vibrating ions in finite temperature crystals), this will lead to general (interaction independent) {\it non-relativistic bounds on single particle speeds} \cite{planckianAOP}. These are the sort of bounds that we will focus on in this work.

\section{Diffusion bounds from uncertainty relations}
\label{sec:uncertainty_sec}
To illustrate how the inequalities of the previous Section may carry implications for transport properties, we first discuss the diffusion constant. With $v_{i \ell}$ denoting the $\ell$-th Cartesian component of the velcoity of the $i$-th particle, the Green-Kubo relation \cite{Green1954,KMS2} expresses the diffusion constant as an integral of a single particle velocity autocorrelation function \cite{liqGv1,liqGv2,liqGv3,liqGv4,liqGv5,liqGv6,liqGv7,liqGv8}, 
\begin{eqnarray}
\label{eq:diffusion}
D =  \int_{0}^{\infty} dt ~ {\sf{Tr}} (\rho_{\Lambda}^{\sf canonical}~ v_{i \ell}^{H}(t)~ v_{i \ell}^{H}(0)) \nonumber
\\ \equiv \int_{0}^{\infty} dt~ G_{v}(t),
\end{eqnarray}
with the canonical probability density $\rho_{\Lambda}^{\sf canonical}=e^{-\beta H_{\Lambda}}/Z_{\Lambda}$ (prior to its replacement by its above discussed classical counterpart $\rho_{\Lambda}^{\sf classical~canonical}$). Starting from its $t=0$ value of the equilibrium average $G_{v}(0)=\langle v_{i \ell}^{2} \rangle$, the velocity autocorrelation function $G_v(t)$ remains positive at sufficiently short times $0\le t <t^{v}$. Here, we define $t^v$ to be the first zero of $G_{v}(t)$ if any such finite time zero of the autocorrelation function exists (i.e.,  $G_v(t=t^v) =0)$ and otherwise will set $t_{v} = \infty$. In some circles, the time interval of negative $G_v(t)$ is called the ``correlation hole'' \cite{Lev86}. In systems such as dilute gases, $G_v(t)$ decays exponentially, remaining positive semi-definite (vanishing only as $t \to \infty$) and no such region exists. However, in solids, liquids, and dense gases, $G_v(t)$ can assume negative values, exhibiting oscillations driven by restoring forces and caging effects \cite{liqGv1,liqGv2,liqGv3,liqGv4,liqGv5,liqGv6,liqGv7,liqGv8}. While these oscillations might  contribute little to the integral of Eq.~\eqref{eq:diffusion} in fluids, they may dominate in solids. In the presence of sufficiently strong disorder, the diffusion constant may vanish, $D=0$, as in Anderson localized systems  \cite{Anderson,NFL}. Sign changes of $G_{v}(t)$ also arise in gases under external magnetic fields due to cyclotron motion. The diffusion constant is equal to the asymptotic $t \to \infty$  limit of the derivative of the squared total displacement during time $t$, i.e., $D = (\lim_{t \to \infty} \partial_{t} \langle (\Delta r^{H}_{il}(t))^{2} \rangle)/2$ where $\Delta r^{H}_{il}(t) \equiv (r^{H}_{il}(t) - r^{H}_{il}(0))$ \footnote{Since the squared displacement 
$\langle  (\Delta r^{H}_{il}(t))^{2} \rangle =\langle  (\int_{0}^{t} v^{H}_{il}(t') dt)^2 \rangle = \int_{0}^{t} dt' \int_{0}^{t} dt'' \langle v^{H}_{il}(t') v^{H}_{il}(t'') \rangle = 2 \int_{0}^{t} dt' \int_{0}^{t'}  dt''   \langle v^{H}_{il}(t') v^{H}_{il}(t'') \rangle$, it follows that $\frac{\partial  \langle (\Delta r^{H}_{il}(t))^{2} \rangle }{\partial t} = 2 \int_{0}^{t}  dt'' G_{v}(t'')$.}. In what follows, we will not bound the diffusion constant but rather 
\begin{eqnarray}
\label{eqnD+}
D_{+} \equiv  \int_{0}^{t^v} dt~ G_{v}(t).
\end{eqnarray}
Clearly, if $G_v >0$ for all finite times $t$ then $D_{+}$ will equal the diffusion constant, $D_{+} = D$.
If single particle displacements can be measured, then $D_{+}$ can, in principle, be experimentally determined for general finite $t^v$. This is so since $D_{+} = \partial_t (\langle \Delta r^{H}_{il}(t))^{2} \rangle/2$ when this derivative is evaluated at $t=t^{v}$.
A simple generalization of this time derivative of squared displacement relation holds for times $t_s<t^{v}$ with the integral of $G_v$  in Eq. (\ref{eqnD+}) performed up to time $t_s$ instead of $t^{v}$. To bound $D_{+}$, note that for $0\le t \le t^v$,
\begin{eqnarray}
\label{Gvt}
  G_v(t)
  \ge \max 
  \left\{
  G_{v}(0)-t\left|
  \frac{dG_v}{dt}
  \right|_{\max},0\right\}
\end{eqnarray}
where $\left|
  \frac{dG_v}{dt}
  \right|_{\max}$ is the maximum absolute value of the autocorrelation function time derivative for $t \in [0,t^v]$. 
The integral of Eq. (\ref{eqnD+}) is thus bounded from below, 
\begin{eqnarray}
\label{D+min}
D_{+} \ge \frac{ \big(G_{v}(0)\big)^2}{2\Big|\frac{dG_v}{dt}\Big|_{\max}}.
\end{eqnarray} 
The ratio appearing on the righthand side of this inequality is the area of a triangle whose height is $G_{v}(0)$ and has a base of value $t_{\min}$- the shortest possible time for $G_{v}(t)$ (that is consistent with the local  uncertainty relations) to drop to zero. The single particle momentum component $p^{H}_{i \ell}(t)$ (and hence velocity $v^{H}_{i \ell}(t)$) commutes with all terms in the Hamiltonian of Eq. (\ref{eq:HLAMBDA}) except for the sum of the interaction terms, $\tilde{H}_{i}^{H}(t)= V^{H}_{i}(t) \equiv \sum_{j} {\cal V}^H_{ij}(t)$, that involve the spatial coordinates of particle $i$ and thus contribute to the time derivative of $(v_{i \ell}^{H}(t) v^H_{i \ell}(0))$ in Heisenberg's equation of motion. 
With this non-commuting component of the total Hamiltonian ($H_{i}^{H}(t) \subset H_{\Lambda}$) identified, the recipe of Section \ref{nut} may now be invoked. The local quantum mechanical time-energy uncertainty relations bound how rapidly the autocorrelation function may vary with time (with the latter found to scale as Eq. (\ref{eq:time})). Towards this end, we may employ the local uncertainty relations of Section \ref{Sec:local} with $A= V^{H}_{i}$ and $B = (v_{i \ell}^{H}(t) v_{i \ell}^{H}(0))$
to rigorously bound $|\frac{dG_v}{dt}|$ from above and subsequently obtain an approximate ``classical'' form by evaluating expectation values with the classical canonical probability density. This yields \cite{planckianAOP},  
\onecolumngrid
\begin{equation}
\label{minDiff+}
D_{+} \ge \frac{ \hbar k_{B} T}{4\sqrt{3}~m~\sqrt[4]{{\sf
  Tr}\Big(\rho_{\Lambda}^{\sf classical~ canonical} \big(\Delta
  V_{i}^{H}\big)^4\Big)}}
     \max\left\{
  \frac
  {\sqrt[4]{{\sf Tr}\left(
  \rho_{\Lambda}^{\sf classical~canonical}
  \big(\Delta V_i^{H}\big)^4
  \right)}}
  {\sqrt{{\sf Tr}\left(
  \rho_{\Lambda}^{\sf classical~canonical}
  \big(\Delta V_i^{H}\big)^2
  \right)}}
  ,\sqrt[4]{3}
     \right\},
\end{equation}
\twocolumngrid
with $\Delta 
  V_{i}^{H} \equiv 
  (V_{i}^{H}- \langle V_{i}^{H} \rangle)$. 
  Eq. (\ref{minDiff+}) implies that a sufficiently large viscosity necessarily leads to a breakdown of the Stokes-Einstein relation. As mandated by dimensional analysis and seen by trivial integration  substitutions when computing the classical moments in Eq. (\ref{minDiff+}), for power law interactions, 
  ${\sf{Tr}}\left(\rho_{\Lambda}^{\sf {classical~canonical }}\left( \Delta V_i^H\right)^{p}\right)$ scales as $(k_{B} T)^p$ \cite{planckianAOP}. For such interactions, in the above limit in which the thermal averages are classical,
  \begin{eqnarray}
  \label{d+}
      D_{+} \ge {\cal{O}}(\frac{\hbar}{m}).
  \end{eqnarray}
  Naturally, a diffusion constant scale of $(\hbar/m)$ is set by dimensional considerations yet the value that the ratio between $D_{+}$ (and the diffusion constant $D$) and $(\hbar/m)$ assumes is not necessarily as trivial. In later Sections, we will compare the bound of Eq. (\ref{d+}) on $D_+$ (as well as its comparison to the diffusion constant $D$) with experimental data. In inequalities for the diffusion constant $D$ that we turn to next, the prefactor multiplying $(\hbar/m)$ is $1/(2 \pi)$ (instead of the numerically similar (yet slightly larger) factor of $1/(4 \sqrt[4]{3})$ in Eq. (\ref{minDiff+})). These latter bounds for the diffusion constant {\it are not far} from the actual minimal values of the experimentally measured diffusion constant in some ``classical'' systems such as {\it water at atmospheric pressure}. In general, we will find these bounds to be quite stringently {\it satisfied} for the diffusion constants of {\it all examined gases} while violations may arise in dense fluids (in which, $G_v(t)$
may become negative and thus $|D - D_{+}|$ can become appreciable). The diffusion constant is far smaller than $\hbar/m$ in cryogenic hydrogen and (even less surprisingly) helium liquids. In these latter fluids, apart from correlation hole effects leading to deviations of $D_{+}$ from $D$, evaluating thermal averages with the classical probability density (i.e., replacing $\rho_{\Lambda}^{\sf {canonical }} \to \rho_{\Lambda}^{\sf {classical~canonical }}$) is also unwarranted.

\begin{figure}
    \centering
    \includegraphics[width=\linewidth]{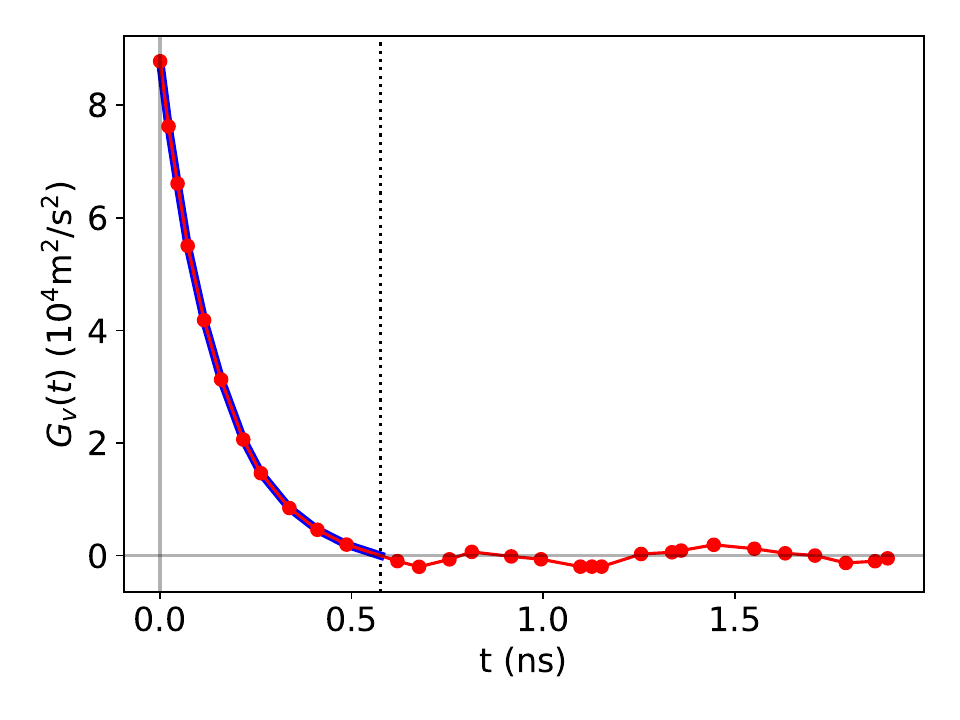}
    \caption{Velocity autocorrelation function of $N_2$ molecules confined within a carbon nanotube \cite{n2inCNT}. 
    For this system, the diffusion constant 
    $D=1.236\times10^5m^2/s$ while the integral of Eq. (\ref{eqnD+}) yields 
    $D_+=1.265\times10^5m^2/s$. The relative difference $(D_{+} - D)/D = 2.3\%$.}
    \label{fig:n2inCNT}
\end{figure}

\begin{figure}
    \centering
    \includegraphics[width=\linewidth]{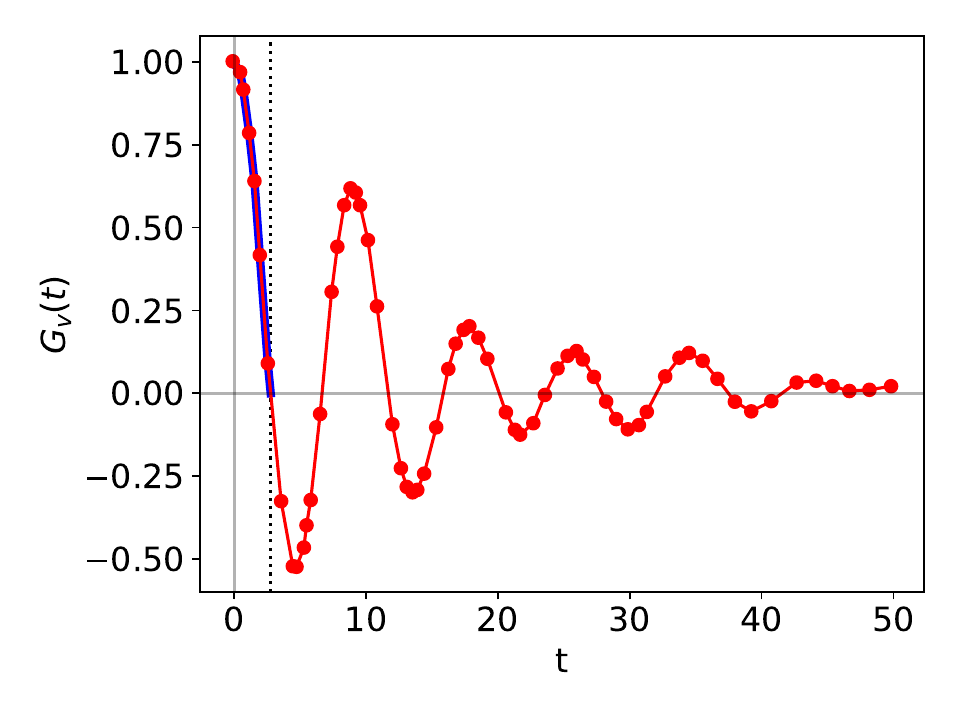}
    \caption{Velocity autocorrelation of charged particles in two dimensions
    interacting via Yukawa interactions and under the influence of a magnetic field \cite{veloAutoCorrMagField}.
    All the quantities are in reduced simulation units. Here, $D=2.386$ while $D_+=1.829$ corresponding to a relative difference of $-23\%$.}
    \label{fig:veloAutoCorrMagField}
\end{figure}

\begin{figure}
    \centering
    \includegraphics[width=\linewidth]{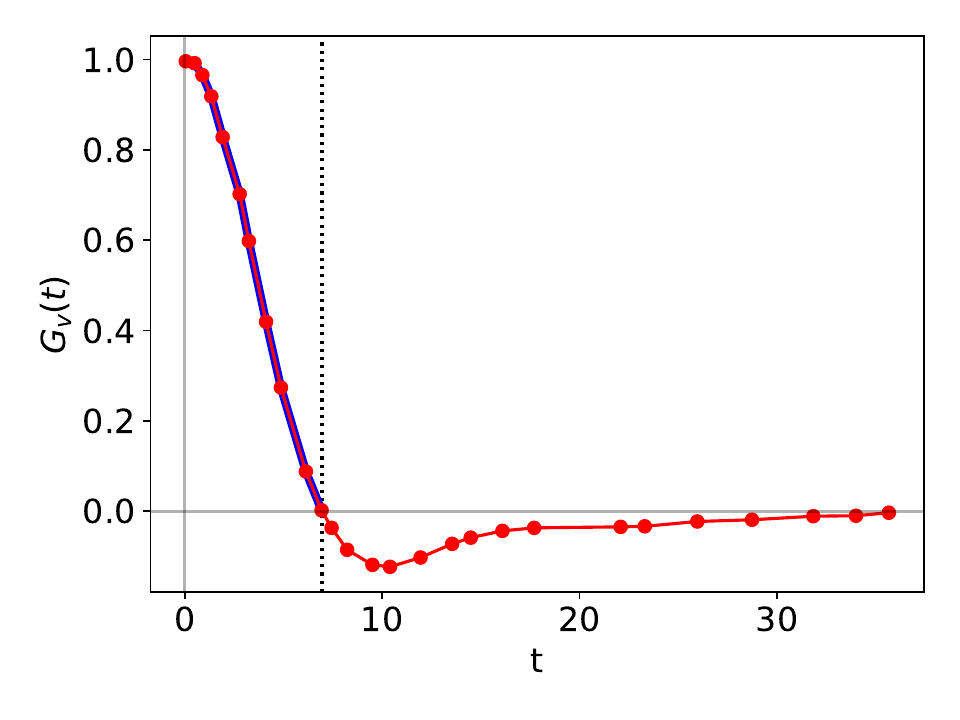}
    \caption{Velocity autocorrelation of particles interacting via a Lennard-Jones potential at $\rho=0.85$ and $T=0.76$ \cite{veloAutocorrLJ} in reduced units. The diffusion constant $D=2.447$ with the integral of Eq. (\ref{eqnD+}) being $D_+=3.679$ corresponding to a relative difference of $+50\%$.}
    \label{fig:veloAutoLJ}
\end{figure}

To better appreciate the difference (and similarity) between $D_{+}$ across various systems, we provide in Figs. \ref{fig:n2inCNT}, \ref{fig:veloAutoCorrMagField} and \ref{fig:veloAutoLJ}, a comparison between these two quantities in certain fluids. 
These and similar other comparisons illustrate that the diffusion constant estimate $D_+$ is generally comparable to the actual diffusion constant, $D$. 
Generally, mismatch between the values of $D_+$ and $D$ becomes more pronounced only under extreme conditions, such as high densities, high pressures, or low temperatures. Deviations may indeed become larger under thees conditions when particle motion is more hindered and $G_v(t)$ may exhibit increasing oscillations. 

\section{Transport bounds from chaos}
\label{sec:chaos}

Extending earlier notions concerning chaos in semiclassical systems \cite{Larkin}, Maldacena {\it et al.} \cite{juan-martin} proposed universal bounds on Lyapunov exponents in thermal quantum systems,
\begin{eqnarray}
\label{eQn:lam} 
\lambda_L \le \frac{2 \pi k_B T}{\hbar},
\end{eqnarray}
as probed by quantum Out-of-Time-Order Correlation functions (OTOCs) \cite{juan-martin,RSL,Larkin}. Insightful   results \cite{juan-martin,subleading,saso2} such as Eq. (\ref{eQn:lam}) have been derived when making relatively modest assumptions. These bounds indeed appear in theories of black holes \cite{juan-martin,op-scram1,Shenker}, the Sachdev-Ye-Kitaev model \cite{SY-SYK,K-SYK,Juan-SYK,0-SYK,1-SYK,2-SYK,3-SYK,4-SYK,5-SYK}, and systems without quasiparticles \cite{PatelSachdev}, which may rapidly thermalize. Connections also emerge between chaotic dynamics, information scrambling \cite{op-scram1,op-scram2,Zanardi}, the Eigenstate Thermalization Hypothesis \cite{Foini,MM19}, and free particle propagation in diverse geometries \cite{Kurchan}. Building on the ``chaos bounds'' of Eq. (\ref{eQn:lam}), we next motivate \cite{planckianAOP} bounds on the diffusion constants similar to those derived in Section \ref{sec:uncertainty_sec}. In a broader context, this reinforces conjectured links between chaos and transport such as those in studies of electronic systems \cite{kapitulnik}. Towards this end, we return to the Green-Kubo relations for semiclassical systems in which, at short times, the autocorrelation functions $G_v(t)$ decay is bounded as
\begin{equation}
\label{Eq:Dexp}
 G_v(t) \gtrsim G_v(0) e^{-t/t_{d}},   
\end{equation}
with a ``dissipation time'' $t_d = 1/\lambda_L$ \cite{juan-martin} where the Lyanpunov exponent $\lambda_L$ satisfies Eq. (\ref{eQn:lam}). More precisely, in quantum chaotic systems, a hierarchy of time scales has been conjectured. The asymptotic form of the most rapid of the decays associated with these (i.e., that dominant at short times) has a characteristic time scale $t_d$ \cite{juan-martin}. For such systems, with the stated assumption of Eq. (\ref{Eq:Dexp}) \cite{juan-martin} with 
the classical dissipation time being the reciprocal of the Lyapunov exponent, we note the Green-Kubo relation (Eq. (\ref{eq:diffusion})) for the diffusion constant trivially leads to the inequality 
\begin{eqnarray}
\label{|GK|}
 D \gtrsim  && \int_{0}^{\infty} dt  ~ e^{-  \lambda_L t}~   \Big\langle \Big( v^{H}(0) \Big)^2  \Big \rangle \nonumber
\\ &&= \frac{1}{\lambda_{L}}  ~   \Big\langle (v^{H}(0))^2 \Big\rangle.
\end{eqnarray}
For Hamiltonians of the form of Eq. (\ref{eq:HLAMBDA}), the {\it classical thermal average}  
$ \Big\langle v_{i \ell}^2 \Big\rangle
=  (k_{B} T/m)$. This implies that when thermal averages become classical, the diffusion constant obeys a simple universal inequality, 
\begin{eqnarray}
\label{DOTOC}
\boxed{D  \gtrsim \frac{\hbar}{ 2 \pi m}.}
\end{eqnarray} 
We underscore that the derivation of this bound relied on extending the short time inequality of Eq. (\ref{Eq:Dexp}) to all times. This focus on short time contributions is akin to that in Section \ref{sec:uncertainty_sec}  when computing $D_{+}$ as defined by the short time integral of Eq. (\ref{eqnD+}). Although, in general, it is less tight than Eq. (\ref{minDiff+}), an advantage of the inequality of Eq. (\ref{DOTOC}) is that it is {\it independent of the specific form of the interactions} ${\cal{V}}$. Further yet, since (i) the bound of Eq. (\ref{eQn:lam}) is argued to be universal \cite{juan-martin} in describing the maximal relaxation rate with $t_d = 1/\lambda_{L}$ and as (ii) the Green-Kubo relations express {\it any} response function 
$\gamma = \int_{0}^{\infty} dt ~   \langle \dot{Y}^{H}(0) \dot{Y}^{H}(t) \rangle$ 
as the time integral of a respective autocorrelation function (being that of the single particle velocities for the diffusion constant), it follows that if short time contributions dominate in these integrals, an analog of Eq. (\ref{DOTOC}) applies to all transport coefficients. That is,
\begin{eqnarray}
\label{|GK|++}
\boxed{\gamma  \gtrsim \frac{\hbar}{2 \pi k_{B} T}  \Big\langle \Big(\dot{Y}^{H}(0) \Big)^2  \Big \rangle.
}
\end{eqnarray}
As before, when the system becomes ``classical'', we may equate the equal time equilibrium expectation value 
of the relevant local quantity $ \langle (\dot{Y}^{H}(0) )^2  \rangle$ to its pertinent classical thermal average. 

In Ref. \cite{planckianAOP}, bounds that are {\it tighter} than those of Eq. (\ref{eQn:lam}) were derived for the rates of change of various local quantities using the recipe of Sections \ref{nut} and \ref{Sec:local} to the difference between two identical observables in, respectively, two decoupled systems that evolved with similar Hamiltonians yet different initial conditions. These Lyapunov exponent bounds did not need the introduction of OTOCs. For instance, assuming Eq, (\ref{Eq:Dexp}), for the diffusion constant (for which $Y^H_i$ is the local particle displacement), we found that $ D \gtrsim \frac{\hbar}{2 \sqrt{d} m}$  (larger by a factor of $\pi/\sqrt{3} \sim 1.81$ than the lower bound of Eq. (\ref{DOTOC})). However, unlike Eq. (\ref{eQn:lam}), these bounds generally depended on the local observable $Y^{H}$ examined (as well as, in numerous cases, a dependence on the specific form of the interactions ${\cal{V}}$ that were present).

\section{Viscosity bounds}
Viscosity bounds may be derived similar to those derived for the diffusion using the methods of Sections \ref{sec:uncertainty_sec} and \ref{sec:chaos}. 
These will generally depend on system details  such as the interactions ${\cal V}$. In what follows, we review and discuss Eyring inspired bounds for the viscosity \cite{Eyring1,Eyring2,nnbk,planckianAOP,Anup,zoharViscosityPRR} that lead to a form that depends only on the number density (number of particles per unit volume) ${\sf n}$. Here, we will further part from our more general recipe of Sections \ref{nut} and \ref{Sec:local}. The form of these bounds was inspired (yet treated very differently from his model for ``hole'' motion in liquids \cite{Eyring2} in our own work) by Eyring's studies of chemical reaction rates \cite{Eyring1}.  

What now follows is a review of an intuitive argument for such a bound. As is well known, in conventional ``classical'' systems, the shear viscosity exhibits a  minimum as a function of the temperature $T$ at the lower temperature end of the gaseous phase. This minimum appears since in the liquid the viscosity generally drops with increasing temperature while in the gas the viscosity is monotonically increasing. To obtain a trivial bound, similar to other works, e.g., \cite{kostya2} we examined the viscosity in the gas phase \cite{nnbk,planckianAOP,Reif}. The shear viscosity $\eta$ of a classical gas is related to the collision time $\tau_{\sf coll}$ by 
\begin{equation}
\label{nhT}
    \eta = {\sf n} k_{B} T \tau_{\sf coll}.
\end{equation}
We next write a rather trivial and general bound on the collision time in a classical gas. In classical gases, the mean free path exceeds the de-Broglie wavelength and thus the thermal ratio between de-Broglie wavelength $h/{|\bf{p}}|$ of the particles and their speed $|{\bf{p}}|/m$ is a lower bound on $\tau_{\sf coll}$. We return to the non-relativistic Hamiltonians of Eq. (\ref{eq:HLAMBDA}) having a potential energy depending only on particle position. Regardless of the interaction strengths, particle density, and all other details, the classical thermal average of the latter ratio is readily evaluated being, similar to the theme of this article, a local (here, a single particle) average,  
\begin{equation}
\label{classicalh}
\frac{\int d^{3}{\bf{p}}~~ e^{- \beta {\bf{p}}^{2}/(2m)}~~ (hm/{\bf{{p}}}^2)}{\int d^{3}{\bf{p}}~~  e^{- \beta {\bf{p}}^{2}/(2m)}} = \frac{h}{k_{B} T}.
\end{equation}
Plugging this into Eq. (\ref{nhT}) leads to a trivial viscosity bound \cite{planckianAOP} of a non-degenerate classical gas (and thus, as explained above, of conventional classical gases in general) that is independent of {\it all details apart from the particle density},
\begin{eqnarray}
\label{etanh}
\boxed{\eta \ge {\sf n} h.}
\end{eqnarray}
This bound is similar to more rigorous inequalities derived from the local uncertainty relations \cite{planckianAOP} yet is generally tighter than these. We reiterate and underscore that the average of Eq. (\ref{classicalh}) {\it always} applies in classical systems regardless of interaction details and particle density.  By contrast, Eq. (\ref{nhT}) along with the condition of the mean free path being smaller than the de Broglie wavelength only holds for non degenerate classical gases. As we discuss next the bound of Eq. (\ref{etanh}) may be violated in low temperature helium gases.

\section{Comparison with experimental data}

\begin{figure*}
    \centering
    \includegraphics[width=\textwidth]{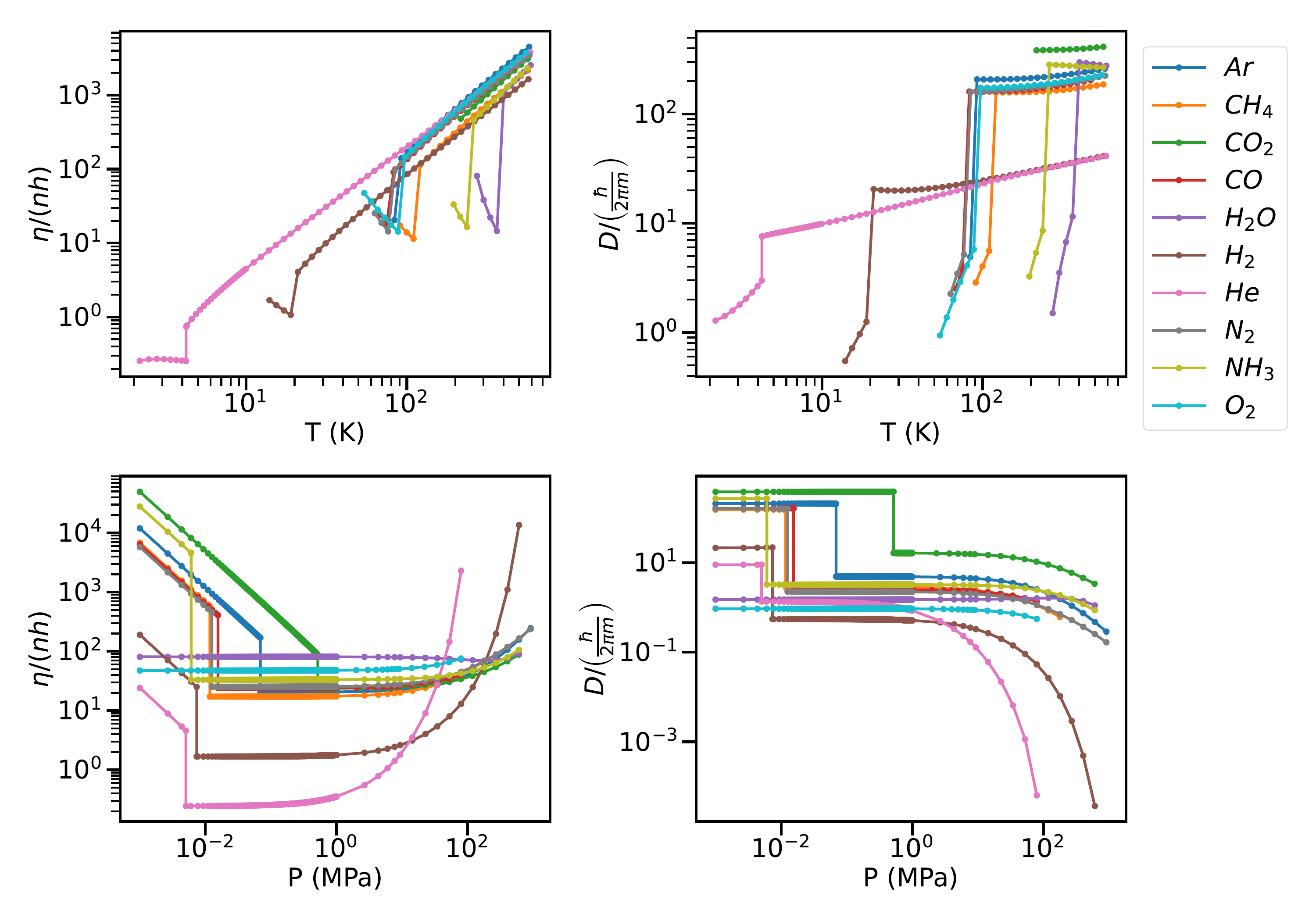}
    \caption{Comparison of the proposed bounds with experimental data \cite{nistThermophysical} for different systems. Left column: Variation of the shear viscosity in units of the bound of Eq.  (\ref{etanh})  
    plotted as a function of temperature and pressure for isobaric and isothermal data respectively.
    Right column: The diffusion constant in units of the bound of Eq. (\ref{DOTOC}) as a function of temperature and pressure for isobaric and isothermal data respectively. The isobaric data is at atmospheric pressure and the isothermal data is at the lowest temperature at which data was available (close to the melting point at atmospheric pressure). With the exception of helium and hydrogen, the bounds are generally satisfied (especially in their plateau like regions).}
    \label{fig:eta-D-TP-variation}
\end{figure*}

We finally turn again to experimental data  \cite{nistThermophysical} and contrast our bounds with these. In calculating the diffusion constant, we use the Stokes-Einstein relation (SER) $D=\frac{k_BT}{6\pi\eta R}$ \cite{SER} relating the diffusion constant to the shear viscosity $\eta$ with $R$ denoting the molecular radius. As is well known, the classical SER \cite{SER} has been observed to fail in numerous experiments   \cite{secor67seViolation,mccall69seViolation,zager82seViolation,rossler90seViolation,Fujara1992}
and {\it in silico} \cite{jungChandler04seViolation,SER2,shiladityaSEviolation,Soklaski}. Particularly well studied is SER violation in the above viscous glassforming systems   \cite{serViolationHodgdonStillinger93} due to ``dynamical heterogeneities'' \cite{DH} (i.e., caused by the presence of itinerant, SER violating, particles that augment more sedentary ones that respect the SER, e.g., \cite{SER1,SER2}). Turning to the systems that we examine in Fig. (\ref{fig:eta-D-TP-variation}) (and Appendix \ref{data}), apart from itineracy due to the notable zero point motion in hydrogen and helium and helium superfluidity, additional departures from classical expectations are further bolstered by numerous other quantum behaviors \cite{SER3}. Whenever it arises, a breakdown of the classical SER will lead to erroneous (lower deduced) values of the diffusion constant from measurements of the viscosity. We suspect that this effect may underlie the more visibly prominent low deduced diffusion constant values in the lower right panel of Fig. \ref{fig:eta-D-TP-variation}. Indeed, in systems having particles of enhanced mobility, the diffusion constant is, generally, far larger \cite{SER1,SER2} than the SER would predict from the measured viscosity. In Appendix \ref{data}, we provide separate tables highlighting the different regimes. In Fig. \ref{fig:eta-D-TP-variation}, we collate the data. Perusing Fig. \ref{fig:eta-D-TP-variation} (and tables in the Appendix), we observe that over the set of examined systems, the bounds of Eqs. (\ref{DOTOC}, \ref{etanh}) appear to be generally satisfied with the (not too surprising) exception of low temperature helium and hydrogen. We reiterate that apart from the obvious invalid use of the classical thermal averaging in these systems, $\rho_{\Lambda}^{\sf {canonical }} \to \rho_{\Lambda}^{\sf {classical~canonical }}$ and attendant SER (especially in cryogenic helium) and possible velocity autocorrelation oscillations (i.e., for which extending short time inequalities such as those of Eq. (\ref{Eq:Dexp}) to long times is invalid). Our bounds are generally more mstrongly obeyed (i.e., the experimental values are substantially larger than our bounds) in the vapor phase where these classical assumptions that we made in arriving at simplified forms may be warranted. The general satisfiability of our viscosity bound bolsters similar trends found across other systems \cite{zoharViscosityPRR,Anup}.

\section{Conclusions}
We outlined how basic ``Planckian'' bounds appear in quantum thermal systems \cite{planckianAOP,macroscopic}. Simplifying our exact local uncertainty relation based inequalties with the aid of classical thermal averages (instead of performing quantum averaging) and assumption of dominant finite time autocorrelation function contributions to the Green-Kubo type integrals \footnote{As discussed in Section \ref{sec:uncertainty_sec}, finite time integrals of the form of $D_{+}$ of Eq. (\ref{eqnD+}) for which we may write down bounds sans the assumption of dominant short time contributions to Green-Kubo relations can, be directly deduced form experimental data when single particle displacements are measurable.} as well as alternate considerations, we illustrated how rather universal bounds on transport coefficients may be derived. Our simplified inequalities appear to be generally satisfied, with the consistent relatively strong exception of cryogenic helium (for both its  shear viscosity and diffusion constant) and hydrogen (for the diffusion constant) when the classical SER is assumed to be applicable to deduce the diffusion constant from measurements of the viscosity. Planckian time scale bounds may carry additional implications for quantum dynamics and minimal time scales required for thermalization (viable effective rapid ``collapse'' of expectation values of local observables to their thermal averages \cite{planckianAOP,macroscopic}). In Appendix \ref{app:disorder}, we briefly discuss possible extensions of our bounds to disordered systems. 

\appendix

\section{Tabulated experimental data comparisons}
\label{data}
 In Tables \ref{tab:Dmin_isobaric_liquid} and \ref{tab:Dmin_isobaric_vapor}, we list the minimal diffusion constants of some fluids at one atmospheric pressure as a function of temperature. In calculating the diffusion constant, we use the SER, $D=\frac{k_BT}{6\pi\eta R}$, relating the diffusion constant $D$ to the shear viscosity $\eta$ with $R$ denoting the molecular radius. For estimating $R$, we use the data for kinetic diameters available in Ref. \cite{breck1973zeolite}. For each system, we separately provide the minimum diffusion constant value measured in the liquid and vapor phases. The temperature at which these minima appear are listed in the last, rightmost, column of the table. In the penultimate column, we write down the ratio of the minimum diffusion constant to our proposed bound of Eq. (\ref{DOTOC}). We see that in most of the rows in the table, our bound is respected and is also relatively tight. However, for liquid hydrogen (and to a lesser extent in oxygen) and our bound is violated. We suspect that this may largely be due to the breakdown of the SER, the inapplicability of classical thermal averaging, and/or larger deviations between $D$ and $D_+$.
\begin{table}[H]
    \centering
    \begin{tabular}{|c|c|c|c|}
\hline
System & $D_{\min}$ ($m^2/s$)& $\displaystyle {D_{\min}}\Big/{\frac{\hbar}{2\pi m}}$ & $T$ ($K$)\\
\hline
$Ar$ & 1.244$\times10^{-9}$ & 4.916 & 83.81\\
$CH_4$ & 1.803$\times10^{-9}$ & 2.861 & 90.69\\
$CO$ & 9.100$\times10^{-10}$ & 2.521 & 68.16\\
$H_2O$ & 8.431$\times10^{-10}$ & 1.503 & 273.2\\
$H_2$ & 2.747$\times10^{-9}$ & 0.5478 & 13.96\\
$He$ & 3.248$\times10^{-9}$ & 1.286 & 2.177\\
$N_2$ & 8.147$\times10^{-10}$ & 2.258 & 63.15\\
$NH_3$ & 1.930$\times10^{-9}$ & 3.251 & 195.5\\
$O_2$ & 2.973$\times10^{-10}$ & 0.9411 & 54.36\\
\hline
    \end{tabular}
    \caption{
    The minimum values of the diffusion constants for isobaric data at one atmospheric pressure  for various {\it liquids}.  
    The second last column gives the minimum diffusion constant in units of our proposed bound of Eq. (\ref{DOTOC}). 
    The last column gives the temperature at which the minimum occurs. 
    The data are from Ref. \cite{nistThermophysical}. Deviations appear for low temperature hydrogen and oxygen possibly due to violations of the SER and/or sufficiently large oscillations of $G_v$ leading to notable differences between $D$ and $D_+$.}. 
    \label{tab:Dmin_isobaric_liquid}
\end{table}
\begin{table}[H]
    \centering
    \begin{tabular}{|c|c|c|c|}
\hline
System & $D_{\min}$ ($m^2/s$)& $\displaystyle {D_{\min}} \Big/{\frac{\hbar}{2\pi m}}$ & $T$ ($K$)\\
\hline
$Ar$ & 5.230$\times10^{-8}$ & 206.7 & 108.8\\
$CH_4$ & 9.934$\times10^{-8}$ & 157.6 & 135.7\\
$CO_2$ & 8.790$\times10^{-8}$ & 382.7 & 216.6\\
$CO$ & 5.793$\times10^{-8}$ & 160.5 & 81.64\\
$H_2O$ & 1.548$\times10^{-7}$ & 275.9 & 600.2\\
$H_2$ & 9.966$\times10^{-8}$ & 19.88 & 27.96\\
$He$ & 1.909$\times10^{-8}$ & 7.560 & 4.224\\
$N_2$ & 5.718$\times10^{-8}$ & 158.5 & 77.36\\
$NH_3$ & 1.577$\times10^{-7}$ & 265.8 & 600.5\\
$O_2$ & 5.490$\times10^{-8}$ & 173.8 & 98.36\\
\hline
    \end{tabular}
    \caption{
    The minimum values of the diffusion constants for isobaric data at one atmospheric pressure for various systems in the {\it vapor phase}.  
    The second last column gives the minimum diffusion constant in units of our proposed bound of Eq. (\ref{DOTOC}). 
    The last column gives the temperature at which the minimum occurs. 
    The data weree obtained from Ref. \cite{nistThermophysical}.
    The diffusion bound of Eq. (\ref{DOTOC}) is satisfied for all examined gases. 
    }
    \label{tab:Dmin_isobaric_vapor}
\end{table}
In Table 
\ref{tab:Dmin_isothermal_vap}, we list the diffusion constants of various systems, again using the viscosity values and the SER, but now at fixed temperature, with varying pressure (between 0.001 and 1000 MPa). 
\begin{table}[H]
    \centering
    \begin{tabular}{|c|c|c|c|c|}
\hline
System & $D_{\min}$ ($m^2/s$)& $\displaystyle {D_{\min}}\Big/{\frac{\hbar}{2\pi m}}$ & $T$ ($K$) & $P$ ($MPa$)\\
\hline
$Ar$ & 5.267$\times10^{-8}$ & 208.2 & 83.81 & 0.06889\\
$CH_4$ & 9.688$\times10^{-8}$ & 153.7 & 90.69 & 0.001\\
$CO_2$ & 8.780$\times10^{-8}$ & 382.3 & 216.6 & 0.001\\
$CO$ & 5.893$\times10^{-8}$ & 163.3 & 68.16 & 0.01554\\
$H_2$ & 1.077$\times10^{-7}$ & 21.49 & 13.96 & 0.001\\
$He$ & 2.279$\times10^{-8}$ & 9.024 & 2.177 & 0.001\\
$N_2$ & 5.807$\times10^{-8}$ & 161.0 & 63.15 & 0.01252\\
$NH_3$ & 1.609$\times10^{-7}$ & 271.1 & 195.5 & 0.001\\
\hline
    \end{tabular}
    \caption{
    The minimum values of the diffusion constants for various systems in their {\it vapor} phase using isothermal data at the minimum temperature at which data was available for each system (close to the melting temperature at one atmospheric pressure). 
    The third last column gives the minimum diffusion constant in units of our proposed bound of Eq. (\ref{DOTOC}). 
    The last two columns give the temperature of the isothermal dataset and pressure at which the minimum occurs. 
    The pressure was varied between 0.001 and 1000 $MPa$.
    The data are from Ref. \cite{nistThermophysical}.}
    \label{tab:Dmin_isothermal_vap}
\end{table}

Summarizing the above comparisons, the diffusion constant bound of Eq. (\ref{DOTOC}) is {\it satisfied for all examined gases}. {\it In liquids, deviations appear}. These may originate from the neglect of sign oscillations of the velocity autocorrelation functions in liquids (i.e., a difference between $D_+$ and $D$) and/or violations of the Stokes-Einstein relation.

We next broadly contrast our viscosity bounds of Eq. (\ref{etanh}) with experimental data \cite{nistThermophysical}. In Tables \ref{tab:etamin_isobaric_liq} and \ref{tab:etamin_isothermal_liq}, we tabulate isobaric and isothermal data, respectively, for the minimum values of the coefficient of viscosity in the liquid phase, as was done earlier for the diffusion constant. In Tables \ref{tab:etamin_isobaric_vap} and \ref{tab:etamin_isothermal_vap}, we do the same for the vapor phase. There are some violations of our bound at extreme conditions such as low temperature or high pressure.
For isobaric data at one atmospheric pressure, violations are seen for,
liquid hydrogen at 20.37K (marginal violation), and,
both liquid and vapor helium at 4.224K.
For the isothermal dataset, our bound is violated for liquid helium at 2.177K.
\begin{table}[H]
    \centering
    \begin{tabular}{|c|c|c|c|}
\hline
System & $\eta_{\min}$ ($\mu Pa\cdot s$)& $\displaystyle {\eta_{\min}}\big/{\sf n}h$ & $T$ ($K$)\\
\hline
$Ar$ & 260.3 & 18.67 & 87.3\\
$CH_4$ & 116.8 & 11.12 & 111.7\\
$CO$ & 165.4 & 14.64 & 81.64\\
$H_2O$ & 281.7 & 13.27 & 373.1\\
$H_2$ & 13.49 & 0.9619 & 20.37\\
$He$ & 3.155 & 0.2539 & 4.224\\
$N_2$ & 160.7 & 13.99 & 77.36\\
$NH_3$ & 251.7 & 15.76 & 239.8\\
$O_2$ & 194.7 & 13.68 & 90.19\\
\hline
    \end{tabular}
    \caption{
    The minimum shear viscosity values of various {\it liquids} using isobaric data at atmospheric pressure.
    The second last column gives the minimum coefficient of viscosity in units of our proposed bound of Eq. (\ref{etanh}). 
    The last column gives the temperature at which the minimum occurs. 
    The data are from Ref. \cite{nistThermophysical}.}
    \label{tab:etamin_isobaric_liq}
\end{table}

\begin{table}[H]
    \centering
    \begin{tabular}{|c|c|c|c|c|}
\hline
System & $\eta_{\min}$ ($\mu Pa\cdot s$)& $\displaystyle {\eta_{\min}}\big/{\sf n}h$ & $T$ ($K$) & $P$ ($MPa$)\\
\hline
$Ar$ & 290.2 & 20.50 & 83.81 & 0.06889\\
$CH_4$ & 193.6 & 17.24 & 90.69 & 0.0117\\
$CO_2$ & 253.4 & 23.71 & 216.6 & 0.518\\
$CO$ & 291.5 & 24.09 & 68.16 & 0.01554\\
$H_2O$ & 1650 & 70.16 & 273.2 & 141.7\\
$H_2$ & 25.59 & 1.679 & 13.96 & 0.007358\\
$He$ & 3.573 & 0.2455 & 2.177 & 0.005039\\
$N_2$ & 311.6 & 25.22 & 63.15 & 0.01252\\
$NH_3$ & 570.6 & 33.19 & 195.5 & 0.006053\\
$O_2$ & 773.6 & 47.50 & 54.36 & 0.001\\
\hline
    \end{tabular}
    \caption{
    The minimum values of the coefficients of viscosity for various liquids using isothermal data at the minimum temperature at which data was available for each system. 
    The third last column gives the minimum coefficient of viscosity in units of our proposed bound of Eq. (\ref{etanh}). The last two columns give the temperature of the isothermal dataset and pressure at which the minimum occurs. 
    The pressure was varied between 0.001 and 1000 $MPa$.
    The experimental data are from Ref. \cite{nistThermophysical}.}
    \label{tab:etamin_isothermal_liq}
\end{table}

\begin{table}[H]
    \centering
    \begin{tabular}{|c|c|c|c|}
\hline
System & $\eta_{\min}$ ($\mu Pa\cdot s$)& $\displaystyle {\eta_{\min}}\big/{\sf n}h$ & $T$ ($K$)\\
\hline
$Ar$ & 7.169 & 124.3 & 87.3\\
$CH_4$ & 4.324 & 95.71 & 111.7\\
$CO_2$ & 10.94 & 480.0 & 216.6\\
$CO$ & 5.490 & 88.37 & 81.64\\
$H_2O$ & 12.23 & 923.9 & 373.1\\
$H_2$ & 0.9963 & 3.778 & 20.37\\
$He$ & 1.246 & 0.7397 & 4.224\\
$N_2$ & 5.444 & 82.87 & 77.36\\
$NH_3$ & 8.066 & 386.8 & 239.8\\
$O_2$ & 6.950 & 124.8 & 90.19\\
\hline
    \end{tabular}
    \caption{
    The minimum values of the coefficients of viscosity for various systems in the {\it vapor} phase, using isobaric data at one atmospheric pressure.
    The second last column gives the minimum coefficient of viscosity in units of our bound of Eq. (\ref{etanh}).
    The last column gives the temperature at which the minimum occurs. 
    The data are from Ref. \cite{nistThermophysical}.}
    \label{tab:etamin_isobaric_vap}
\end{table}

\begin{table}[H]
    \centering
    \begin{tabular}{|c|c|c|c|c|}
\hline
System & $\eta_{\min}$ ($\mu Pa\cdot s$)& $\displaystyle {\eta_{\min}}\big/{\sf n}h$ & $T$ ($K$) & $P$ ($MPa$)\\
\hline
$Ar$ & 6.824 & 1.191$\times10^4$ & 83.81 & 0.001\\
    $CH_4$ & 3.598 & 576.9 & 90.69 & 0.0117\\
$CO_2$ & 10.89 & 87.30 & 216.6 & 0.518\\
$CO$ & 4.492 & 6375 & 68.16 & 0.001\\
$H_2$ & 0.6452 & 25.10 & 13.96 & 0.007358\\
$He$ & 0.5376 & 4.592 & 2.177 & 0.005039\\
$N_2$ & 4.365 & 5740 & 63.15 & 0.001\\
$NH_3$ & 6.838 & 4581 & 195.5 & 0.006053\\
\hline
    \end{tabular}
    \caption{
    The minimum values of the coefficients of viscosity for various systems in the {\it vapor} phase using isothermal data at the minimum temperature at which data was available for each system. 
    The third last column gives the minimum coefficient of viscosity in units of our proposed bound of Eq. (\ref{etanh}). The last two columns give the temperature of the isothermal dataset and pressure at which the minimum occurs. 
    The pressure was varied between 0.001 and 1000 $MPa$.
    The data are from Ref. \cite{nistThermophysical}.}
    \label{tab:etamin_isothermal_vap}
\end{table}
In Tables \ref{tab:etaByNh_isobaric_liq} and \ref{tab:etaByNh_isothermal_liq}, we tabulate the ratio  $\displaystyle\frac{\eta}{{\sf n}h}$ in the liquid phase under isobaric and isothermal conditions, respectively. 
In Tables \ref{tab:etaByNh_isobaric_vap} and \ref{tab:etaByNh_isothermal_vap}, we do the same for the vapor phase.
For this quantity, the bounds are violated for the same instances as in the viscosity data in the previous two tables.
\begin{table}[H]
    \centering
    \begin{tabular}{|c|c|c|c|}
\hline
System & $\displaystyle \left(\frac{\eta}{{\sf n}h}\right)_{\min}$ & $T$ ($K$)\\
\hline
$Ar$ & 18.67 &  87.3\\
$CH_4$ & 11.12 &  111.7\\
$CO$ & 14.64 &  81.64\\
$H_2O$ & 13.27 &  373.1\\
$H_2$ & 0.9619 &  20.37\\
$He$ & 0.2539 &  4.224\\
$N_2$ & 13.99 &  77.36\\
$NH_3$ & 15.76 &  239.8\\
$O_2$ & 13.68 &  90.19\\
\hline
    \end{tabular}
    \caption{
    The minimum values of the ratio of the coefficient of viscosity to the product of the number density and Planck's constant for various {\it liquids} using isobaric data at one atmospheric pressure. 
    The last column gives the temperature at which the minimum occurs. 
    The data were obtained from Ref. \cite{nistThermophysical}.}
    \label{tab:etaByNh_isobaric_liq}
\end{table}

\begin{table}[H]
    \centering
    \begin{tabular}{|c|c|c|c|}
\hline
System & $\displaystyle \left(\frac{\eta}{{\sf n}h}\right)_{\min}$ & $T$ ($K$) & $P$ ($MPa$)\\
\hline
$Ar$ & 20.5 &  83.81 & 0.06889\\
$CH_4$ & 17.24 &  90.69 & 0.0117\\
$CO_2$ & 23.71 &  216.6 & 0.518\\
$CO$ & 24.09 &  68.16 & 0.01554\\
$H_2O$ & 69.54 &  273.2 & 195\\
$H_2$ & 1.679 &  13.96 & 0.007358\\
$He$ & 0.2455 &  2.177 & 0.005039\\
$N_2$ & 25.22 &  63.15 & 0.01252\\
$NH_3$ & 33.19 &  195.5 & 0.006053\\
$O_2$ & 47.5 &  54.36 & 0.001\\
\hline
    \end{tabular}
    \caption{
    The minimum values of the ratio of the coefficient of viscosity to the product of the number density and Planck's constant for isothermal data for various {\it liquids} at the minimum temperature at which data was available for each system (close to the melting temperature at one atmospheric pressure).
    The third last column gives the minimum ratio of the coefficient of viscosity to the product of the number density and Planck's constant in units of our proposed bound of Eq. (\ref{etanh}).
    The last two columns give the temperature of the isothermal dataset and pressure at which the minimum occurs. 
    The pressure was varied between 0.001 and 1000 $MPa$.
    The data are from Ref. \cite{nistThermophysical}.}
    \label{tab:etaByNh_isothermal_liq}
\end{table}

\begin{table}[H]
    \centering
    \begin{tabular}{|c|c|c|c|}
\hline
System & $\displaystyle \left(\frac{\eta}{{\sf n}h}\right)_{\min}$ & $T$ ($K$)\\
\hline
$Ar$ & 124.3 &  87.3\\
$CH_4$ & 95.71 &  111.7\\
$CO_2$ & 480 &  216.6\\
$CO$ & 88.37 &  81.64\\
$H_2O$ & 923.9 &  373.1\\
$H_2$ & 3.778 &  20.37\\
$He$ & 0.7397 &  4.224\\
$N_2$ & 82.87 &  77.36\\
$NH_3$ & 386.8 &  239.8\\
$O_2$ & 124.8 &  90.19\\
\hline
    \end{tabular}
    \caption{
    The minimum values of the ratio of the coefficient of viscosity to the product of the number density and Planck's constant for various systems in the {\it vapor} phase using isobaric data at one atmospheric pressure. 
    The last column gives the temperature at which the minimum occurs. 
    Data are from Ref. \cite{nistThermophysical}.}
    \label{tab:etaByNh_isobaric_vap}
\end{table}

\begin{table}[H]
    \centering
    \begin{tabular}{|c|c|c|c|}
\hline
System & $\displaystyle \left(\frac{\eta}{{\sf n}h}\right)_{\min}$ & $T$ ($K$) & $P$ ($MPa$)\\
\hline
$Ar$ & 169.3 &  83.81 & 0.06889\\
$CH_4$ & 576.9 &  90.69 & 0.0117\\
$CO_2$ & 87.3 &  216.6 & 0.518\\
$CO$ & 407.6 &  68.16 & 0.01554\\
$H_2$ & 25.10 &  13.96 & 0.007358\\
$He$ & 4.592 &  2.177 & 0.005039\\
$N_2$ & 455.6 &  63.15 & 0.01252\\
$NH_3$ & 4581 &  195.5 & 0.006053\\\hline
    \end{tabular}
    \caption{
    The minimum values of the ratio of the coefficient of viscosity to the product of the number density and Planck's constant for isothermal data for various systems in the {\it vapor} phase at the minimum temperature at which data was available for each system (close to the melting temperature at one atmospheric pressure).
    The third last column gives the minimum ratio of the coefficient of viscosity to the product of the number density and Planck's constant in units of our proposed bound of Eq. (\ref{etanh}). 
    The last two columns give the temperature of the isothermal dataset and pressure at which the minimum occurs. 
    The pressure was varied between 0.001 and 1000 $MPa$.
    The data are from Ref. \cite{nistThermophysical}.}
    \label{tab:etaByNh_isothermal_vap}
\end{table}
Next (Table \ref{tab:numin}), we focus on another similar quantity, the kinematic viscosity, $\displaystyle \nu\equiv\frac{\eta}{\rho}$ with $\rho$ the mass density. Following Ref. \cite{trachenko2020viscositySciAdv}, we look at isobaric data at the same pressures. Our analysis implies that,
\begin{eqnarray}
    \nu\ge \frac{h}{m},
\end{eqnarray}
$m$, as usual, being the mass of the molecule.
Our bound is violated for hydrogen and helium subject to high pressures.
\begin{table}[H]
    \centering
\begin{tabular}{|c|c|c|c|c|}
\hline
System & $P~(MPa)$ & $\nu_{\min}~(m^2/s)$ & $\nu_{\min}\Big/(h/m)$ & $T~(K)$\\
\hline
$Ar$ & 100 & 7.676$\times10^{-8}$ & 7.684 & 347.6\\
$Ar$ & 20 & 6.005$\times10^{-8}$ & 6.012 & 202.4\\
$CH_4$ & 20 & 1.104$\times10^{-7}$ & 4.439 & 251.5\\
$CO_2$ & 30 & 7.97$\times10^{-8}$ & 8.79 & 399\\
$CO$ & 30 & 8.193$\times10^{-8}$ & 5.749 & 222.5\\
$H_2O$ & 100 & 1.214$\times10^{-7}$ & 5.479 & 706.9\\
$H_2$ & 50 & 1.613$\times10^{-7}$ & 0.8147 & 98.71\\
$He$ & 100 & 7.416$\times10^{-8}$ & 0.7439 & 54.88\\
$He$ & 20 & 5.198$\times10^{-8}$ & 0.5214 & 22.39\\
$N_2$ & 10 & 6.484$\times10^{-8}$ & 4.552 & 156.9\\
$N_2$ & 500 & 1.259$\times10^{-7}$ & 8.838 & 850.9\\
$Ne$ & 300 & 6.936$\times10^{-8}$ & 3.508 & 257.9\\
$Ne$ & 50 & 4.94$\times10^{-8}$ & 2.498 & 102.6\\
$O_2$ & 30 & 6.81$\times10^{-8}$ & 5.461 & 221.6\\
\hline
\end{tabular}
\caption{
    The minimum value, with varying temperature, of the 
    kinematic viscosity, $\nu=\displaystyle \frac{\eta}{\rho}$, for various systems at fixed pressure, chosen from Ref. \cite{trachenko2020viscositySciAdv}. The ratio of $\nu_{\min}$ to our theoretical bound, $h/m$ is also shown. The temperature at which this minimum is seen is given in the last column.}
    \label{tab:numin}
\end{table}

\section{Diffusion in Disordered Systems}
\label{app:disorder}

In illuminating recent work \cite{Heller}, it was found that upon progressively endowing impurities with kinetic energy,  ``Anderson localization ceases to exist'' and, instead, the diffusion constant assumes a finite value $D=\alpha \hbar/m$ with 
$\alpha ={\cal{O}}(1)$. To examine the effect of disorder and general athermal systems, we return to our diffusion constant bounds of Sections \ref{sec:uncertainty_sec} and \ref{sec:chaos} \cite{planckianAOP}. Towards this end, we start with the time integral of Eq. (\ref{eq:diffusion}), $D = \int_{0}^{\infty} dt \langle v^{H}_{i \ell} (t) v^{H}_{i \ell} (0) \rangle = \int_{0}^{\infty} G_{v}(t) dt$, yet now {\it do not} specialize to averages taken with the canonical probability density  $\rho^{\sf canonical}_{\Lambda}$. In the fully localized system of immobile impurities, the diffusion constant vanishes, $D=0$. In the localized regime, the positive and negative valued contributions of the velocity autocorrelation function $G_v(t)$ exactly cancel other when an integration is performed over all times $t$. As the impurity kinetics monotonically increases or as disorder is lowered, the velocity autocorrelation function $G_{v}(t)$ may ultimately transition from featuring pronounced sign fluctuations as a function of time to one in which the velocity autocorrelation contributions from times $0< t<t^v$ of Section \ref{sec:uncertainty_sec} dominate. In this latter case, the minimal value of $D_+$ (Eqs. (\ref{eqnD+}, \ref{Gvt})) is, similar to Eq. (\ref{D+min}), given by   
the area formed by a triangle of height $G_{v}(t=0)$ and width (triangle base) $t_{\min}$ associated with the shortest time $t_{\min}$ possible for $G_{v}(t)$ to decay to zero as determined by our local time-energy uncertainty relations. From these, 
\begin{eqnarray}
t_{\min} \ge \frac{\hbar}{2 \sigma_{{\tilde{H}}_{i}^{H}}}.
\end{eqnarray}
Similar to Section \ref{sec:uncertainty_sec}, ${\tilde{H}}^{H}_{i}(t)$ is the sum ($V_{i}^{H}(t)$) of all potential energy contributions (including, apart from the sum of the pertinent pair interactions ${\cal{V}}$ discussed in Section \ref{sec:uncertainty_sec} now also those of the itinerant external disordered fields) that depend on the coordinates of particle $i$ being the operator that may contribute to the time derivative of $v_{i}^{H}(t)$ and thus of $G_v(t)$. At time $t=0$, in $d$ spatial dimensions, the autcorrelation function is a scaled single particle kinetic energy expectation value, $G_{v}(t=0)=\langle p_{i\ell}^{2}\rangle /m^2 = 2 \langle K_{i} \rangle/(md)$. The lower bound on short time integral of $G_v(t)$ contributions for the diffusion constant then becomes 
\begin{eqnarray}
\label{D+eqeq}
D_{+} \ge 
\frac{\hbar ~~\langle K_{i} \rangle}{8md ~\sigma_{V^H_{i}}}.
\end{eqnarray}

If in the relevant states the kinetic energy $\langle K_i \rangle$ of a given particle $i$ and the standard deviation of its local potential energy $V_i$ are of comparable size, $\frac{\langle T_{i} \rangle}{\sigma_{V_{i}}} \equiv c = {\cal{O}}(1)$ \footnote{When impurities are mobile, this ratio will, in general, be larger than that in the localized system.} then a natural minimal value for the diffusion constant (once contributions of negative $G_v(t)$ no longer cancel those of positive autcorrelation values) may be that of the bound of Eq. (\ref{D+eqeq}), 
$D_{\min} \simeq \alpha \Big(\frac{\hbar}{m} \Big)$ with $\alpha = \frac{c}{8d}$.

\bibliography{bounds_zaanen}
\bibliographystyle{unsrt}
\end{document}